\documentclass[fleqn,twoside]{article}
\usepackage{espcrc2}

\usepackage{graphicx}
\usepackage[figuresright]{rotating}

\newcommand{\AmS}{{\protect\the\textfont2
  A\kern-.1667em\lower.5ex\hbox{M}\kern-.125emS}}

\title{What's Behind Acoustic Peaks in the Cosmic Microwave Background 
Anisotropies}

\author{C. Baccigalupi\address[SISSA]{SISSA/ISAS, Via Beirut 2-4, 34014 Trieste, Italy}, 
A. Balbi\address[ROMA]{Dipartimento di Fisica, Universit\`a di Roma `Tor Vergata' 
and INFN, Sezione di Roma II}, 
S. Matarrese\address{Dipartimento di Fisica `Galileo Galilei',
Universit\`a di Padova, and INFN, Sezione di Padova,\\ 
via Marzolo 8, 35131 Padova, Italy}, 
F. Perrotta \address{Osservatorio Astronomico di Padova, Vicolo
dell'Osservatorio 5, 35122 Padova, Italy}, N. Vittorio\addressmark[ROMA]}

\begin{document}

\begin{abstract}
  We give a brief review of the physics of acoustic oscillations in
  Cosmic Microwave Background (CMB) anisotropies. As an example of the
  impact of their detection in cosmology, we show how the present data
  on CMB angular power spectrum on sub-degree scales can be used to
  constrain dark energy cosmological models.  \vspace{1pc}
\end{abstract}

\maketitle

\section{Introduction}
\label{intro}

As it is well known, see e.g. \cite{PZ}, Cosmic Microwave Background
(CMB) anisotropies can be thought of as fluctuations $\delta T/T$
around the mean black body temperature $T\simeq 2.726$ K of the
cosmological radiation. CMB anisotropies in a particular direction
$\hat{n}$ appear to us as a line of sight integral on the temperature
fluctuations $\delta T/T(\hat{n},z)$ carried by CMB photons last
scattered at a distance $r(z)$ from us, and weighted with the last
scattering probability $P(z)$:
\begin{equation} 
\label{los} 
{\delta T\over T}(here,now,\hat{n})= 
\int_{0}^{\infty}{\delta T\over T}(\hat{n},z)P(z)dz\ .
\end{equation}
The last scattering probability $P(z)$ is fixed by cosmological
recombination history and it turns out to be a narrow peak around a
decoupling redshift $z_{dec}$ with mean and dispersion given by
$1+z_{dec}\simeq 1100\ ,\ \Delta z_{dec}\simeq 100\ ,$ corresponding
to physical distances
\begin{equation}
\label{rdec}
r_{dec}\simeq 6000h^{-1}\ {\rm Mpc}
\ , \Delta r_{dec}\simeq 10 h^{-1}\ {\rm Mpc}\ .
\end{equation}
Therefore, CMB anisotropies can be thought of as a snapshot of the
cosmic thermodynamical temperature in the early Universe.

\begin{figure}[htb]
\vspace{9pt}
\centerline{
\includegraphics[width=20pc,height=20pc]{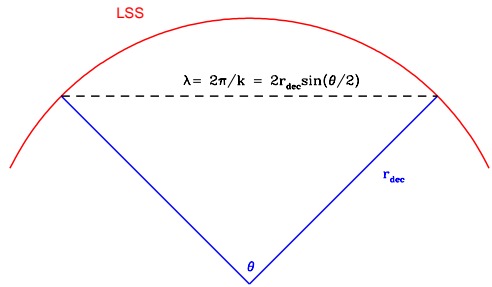}
}
\vskip -.5in
\caption{Angle-scale relation for CMB anisotropy.}
\label{f1}
\end{figure}

Their dependence on the line of sight is usually described through an
expansion into spherical harmonics
\begin{equation}
\label{alm}
{\delta T\over T}(\hat{n}) = \sum_{lm}a_{lm}Y_{lm}(\hat{n})\ .
\end{equation}
The anisotropy two point correlation function, obtained averaging the
product of fluctuations coming from all pairs of directions separated
by an angle $\theta$, can be expanded into Legendre polynomials
$P_{l}(\cos\theta )$:
\begin{eqnarray}
<{\delta T\over T}(\hat{n})
{\delta T\over T}(\hat{n'})>_{\hat{n}\cdot\hat{n'}=\cos\theta}&=&
\nonumber\\
=\sum_{l}{2l+1\over 4\pi}C_{l}P_{l}(\cos\theta )\ &.& 
\label{ctheta}
\end{eqnarray} 
The relation of $C_{l}$s with $a_{lm}$ coefficients is 
\begin{equation}
\label{cl}
C_{l}={1\over 2l+1}\sum_{m}|a_{lm}|^{2}\ .
\end{equation}
At high multipoles, $l\gg 1$, the Legendre polynomials have a sharp
peak at $\theta\simeq 200/l$ degrees; as a consequence, a $C_{l}$
coefficient quantifies the anisotropy power on the same angular scale.
Moreover, taking into account that CMB anisotropies come essentially
from a narrow spherical shell in redshift as in Eq.(\ref{rdec}), also
known as last scattering surface, $C_{l}$ probes perturbations on a
cosmological scale $\lambda$ represented as in figure \ref{f1}.

Given that the cosmological horizon at decoupling subtends roughly one
degree on the sky, this scale separates the sub-horizon from
super-horizon regimes.  After the first discovery of large scale CMB
anisotropies by COBE \cite{COBE}, a breakthrough on the sub-degree
structure of this signal is underway. Data from the two balloon-borne
experiments BOOMERanG and MAXIMA \cite{BOOM,MAX} and the ground based
interferometer DASI \cite{DASI} gave strong evidence of the presence
of a peak at angular scales corresponding to a degree, as well as
important indications for the existence of other peaks on smaller
scales. Forthcoming data from satellites MAP ({\tt
  http://map.nasa.gsfc.gov}) and Planck ({\tt
  http://astro.estec.esa.nl/Planck}), will reveal CMB acoustic
oscillations on the whole sky.

In this paper we give a brief review of the most important physical
mechanisms responsible for the formation of CMB acoustic peaks on
sub-degree angular scales, together with some application of present
data to constrain cosmological models. In Section II we put CMB
anisotropies in the context of cosmological perturbation theory. In
Section III we describe the phenomenology of acoustic peaks.  Finally
in Section IV we show an example of the impact of CMB on cosmology,
briefly describing how these data can constrain cosmologies with dark
energy.

\section{Introducing linear cosmological perturbation theory}
\label{introducing}

Linear cosmological perturbation theory describes small fluctuations
around background quantities in cosmology.  A few years after the
first historical works \cite{PY,B}, a general treatment of
cosmological perturbations has been written \cite{KS} and extended to
the more general context of scalar-tensor theories of gravity
\cite{H}. Recent works \cite{MB,HSWZ} focus on CMB anisotropies,
giving a complete description of their theoretical and
phenomenological aspects. Even if we can give here only the very basic
details of this theory it is useful to put CMB anisotropies into their
context. We restrict our analysis to flat cosmologies.

On cosmological scales, the line element
$ds^{2}=g_{\mu\nu}dx^{\mu}dx^{\nu}$ is described by a perturbed
Friedmann Robertson Walker (FRW) metric tensor
\begin{equation}
\label{gmunu}
g_{\mu\nu}(\vec{x},\eta )=
a(\eta )^{2}[\gamma_{\mu\nu}+h_{\mu\nu}(\vec{x},\eta )]\ , 
\end{equation}
where $a$ is the scale factor describing cosmic expansion and $\eta$
represents the conformal time, defined in terms of the usual time by
$d\eta =dt/a$; $\gamma_{\mu\nu}$ represents he background Minkowski
metric.  The linear fluctuations $h_{\mu\nu}\ll 1$ are conveniently
decomposed into the three different components
\begin{equation}
\label{svt}
h_{\mu\nu}=h_{\mu\nu}^{S}+h_{\mu\nu}^{V}+h_{\mu\nu}^{T}\ ,
\end{equation}
transforming as scalar, vector and tensor quantities under spatial
rotations, respectively \cite{KS}.  Scalar metric fluctuations
represent, in a Newtonian fashion, scalar quantities like a
gravitational potential; vectors modes represent vorticity while
tensors represent cosmological gravitational waves. We restrict our
analysis to scalar perturbations only. Linearity introduces a gauge
freedom so that equations of motion and perturbed quantities can have
different expressions in different frames separated by a linear
coordinate transformation. In other words not all the elements of
$h_{\mu\nu}$ are independent, but some of them can be set to zero via
a proper gauge choice. Here we fix the conformal Newtonian gauge for
which the metric fluctuations appear isotropic with respect to the
cosmic expansion: the non-zero metric perturbations of scalar type are
\begin{equation}
\label{phipsi}
\Psi={h_{00}\over 2}\ ,\ \Phi={h_{11}\over 2}=
{h_{22}\over 2}={h_{33}\over 2}\ .
\end{equation}
On the other hand, linearity allows to analyze cosmological
perturbation evolution in Fourier space, since Fourier modes do not
mix at a linear level. Unless otherwise specified, we write equations
in the Fourier space in the following.

Correspondingly to the metric fluctuations, the stress energy tensor
$T_{\mu\nu}$ is perturbed: for what concerns scalar perturbations, any
cosmological component $x$ admits energy density, velocity and
pressure perturbations due to isotropic and anisotropic stress
\cite{KS}:
\begin{equation}
\label{dxvx}
\delta_{x}\equiv{\delta\rho_{x}\over\rho_{x}}\ ,
\ v_{x}\ ,\ \pi_{L}\ ,\ \pi_{T}\ .
\end{equation}
The above quantities fully describe scalar perturbations for
non-relativistic species, for which velocity perturbations are enough
to describe their peculiar motion with respect to the comoving
expansion. Relativistic species move at the speed of light and are
characterized by a propagation direction $\hat{n}$ which needs to be
properly treated. The dependence on the propagation direction of the
thermodynamical temperature of CMB photons is a key aspect of CMB
perturbation phenomenology. Each Fourier component at wavevector
$\vec{k}$, describing the spatial dependence, is expanded in the
Fourier space, then the dependence on $\hat{n}$ is described through a
spherical harmonic expansion taking the direction in the Fourier
space, $\hat{k}$, as polar axis; for scalar perturbations, only
Legendre polynomials are necessary \cite{HSWZ}:
\begin{equation}
\label{dttl}
{\delta T\over T}(\vec{k},\eta ,\hat{n})=
\sum_{l}\left({\delta T\over T}\right)_{l}(\eta )
P_{l}(\hat{n}\cdot\hat{k})\ .
\end{equation}
Monopole, dipole and quadrupole in the above expansion are related to
density, velocity and stress perturbations of CMB radiation. In
Newtonian gauge these relations, for density and velocity, take the
form
\begin{equation}
\label{dttfluid}
\delta_{\gamma}=4\left({\delta T\over T}\right)_{0}
\ ,\ v_{\gamma}=\left({\delta T\over T}\right)_{1}\ ,
\end{equation}
where the first one recalls the Stephan-Boltzmann law $\rho\propto
T^{4}$. As we expose in the next Section, most of the CMB
phenomenology derives from the behavior of the monopole term.

Unperturbed Einstein equations link the Einstein gravitational tensor
$G_{\mu\nu}$ to the stress energy tensor and describe the scale factor
evolution. Conservation equations $T_{\mu ;\nu}^{\nu}=0$ complete the
evolution system.  In the same way, perturbed Einstein and
conservation equations describe perturbation evolution:
\begin{equation}
\label{dgt}
\delta G_{\mu\nu}=8\pi G\delta T_{\mu\nu}
\ ,\ \delta T_{\mu\, ;\nu}^{\nu}=0\ .
\end{equation}
We do not write here the form of the above equations for all
cosmological species. In the next Section we write only the ones which
are relevant to give a simple understanding of the phenomenology of
CMB anisotropies in terms of the initial conditions which are supposed
to be fixed in the early Universe.

\section{Sub-degree CMB acoustic oscillations}
\label{sub-degree}

The dynamics of the CMB thermodynamical temperature fluctuations 
is dictated by the Thomson scattering cross section. 
For each Fourier mode, evolution equations for each multipole 
defined as in (\ref{dttl}) can be expanded in power of the ratio 
between the wavevector amplitude $k$ and the differential optical 
depth for Thomson scattering $\dot{\tau}$, which corresponds 
to the inverse of the photon mean free path. Dynamics is frozen 
to the initial conditions for scales which are larger than 
cosmological horizon scale $\lambda_{H}$ and the photon mean free path, 
$k\lambda_{H}\ll 1$, $k/\dot{\tau}\ll 1$; 
generically, initial conditions are such that 
only the lowest multipoles are different from zero. 
After the horizon crossing these multipoles evolve giving rise 
to acoustic oscillations but do not transmit power to the higher 
multipoles until decoupling: 
at that time, the mean free path for photons increases 
rapidly up to the cosmological horizon and the oscillations are transmitted 
also to higher multipoles \cite{MB,HSWZ}; therefore, the decoupled photons 
carry the snapshot of acoustic oscillations on sub-horizon scales 
at decoupling. Projected on the last scattering surface, the horizon 
corresponds to a degree in the sky, so that acoustic oscillations 
are mapped by sub-degree CMB anisotropies. 

\begin{figure}[htb]
\vspace{9pt}
\includegraphics[width=20pc,height=18pc]{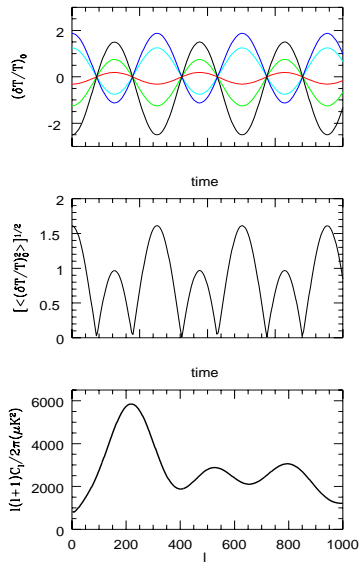}
\caption{Schematical sketch of sub-horizon 
acoustic oscillations for Fourier modes having 
same wavenumber $k$ but different directions 
(top) and their mean quadratic power (middle); 
acoustic peaks in the sky signal in typical 
adiabatic cosmological models (bottom), see text.}
\label{f2}
\end{figure}

For our purposes here, the study of the evolution of 
the zero-point fluctuation $(\delta T/T)_{0}$ is enough. 
At the lowest order in $k/\dot{\tau}$, after horizon crossing, 
neglecting the cosmological friction and the time derivatives 
of gravitational potentials, the zero-point temperature 
fluctuation obeys the simple equation 
\begin{equation}
\label{dtt0}
\ddot{\left({\delta T\over T}\right)}_{0}+
{k^{2}\over 3}\left({\delta T\over T}\right)_{0}=
-{k^{2}\over 3}\Psi\ ,
\end{equation}
which merely represents an harmonic oscillator with radiation 
sound velocity $1/\sqrt{3}$, forced by the cosmological gravitational 
potential $\Psi$ \cite{HSWZ}. In the limit in which the latter is 
constant, the solution, which is 
valid after the horizon crossing time $\eta_{HC}$, 
can be found analytically and describes most of the CMB acoustic 
oscillation phenomenology. 

In the simplest inflationary scenario (see e.g. \cite{PZ}) 
with adiabatic initial conditions, the curvature is perturbed 
with the same power on all cosmological scales at the horizon 
crossing. It can be seen \cite{KS} that curvature 
perturbation is closely related to the cosmological 
gravitational potential $\Psi$. Writing the generic 
Fourier mode as its module multiplied by its phase, 
$\Psi (\vec{k})=|\Psi (\vec{k})|e^{i\phi_{\vec{k}}}$, 
inflationary initial conditions 
generate an initial Gaussian spectrum where the 
first term has in mean the same amplitude for all modes 
at the horizon crossing, while the phase $\phi_{\vec{k}}$ 
is random. 

At the horizon crossing the initial conditions 
for thermodynamical temperature fluctuations are 
simply related to the gravitational potential as 
\begin{equation}
\label{adiaini}
\left({\delta T\over T}\right)_{0}\propto \Psi\ne 0
\ ,\ \dot{\left({\delta T\over T}\right)}_{0}=0\ ,
\end{equation}
so that the solution to (\ref{dtt0}) takes the simple form 
\begin{eqnarray}
\left({\delta T\over T}\right)_{0}(\eta )=
\left[\left({\delta T\over T}\right)_{0}(\eta_{HC})+\Psi\right] &\times& 
\nonumber\\
\times\cos\left[{k(\eta -\eta_{HC})\over\sqrt{3}}\right]-\Psi\ &,&
\label{dtt0adiaaini}
\end{eqnarray}
representing oscillation occurring 
for a given scale $k$; as it is schematically sketched 
in figure \ref{f2}, $\delta T/T$ fluctuations for 
Fourier wavevectors with different direction but same 
amplitude $k$ have random phases 
so that their mean is zero but have the same zeros 
(top panel), and their root mean square power presents 
coherent acoustic peaks (middle panel) located at 
$k(\eta -\eta_{HC})=n\pi\sqrt{3}$, with $n$ integer. 
In the bottom panel, we show the sky signal of a typical 
cosmological model having adiabatic initial conditions. 
The highest peak corresponds to scales crossing the horizon 
just at decoupling, 
and occur at a multipole $l$ corresponding precisely 
to the angle subtended by the horizon at last scattering. 
The second peak at $l\simeq 500$ corresponds to scales 
that were in horizon crossing slightly before decoupling 
and that at the time of decoupling, when the CMB snapshot 
is taken, were in the maximum of their second oscillation. 
In the same way, the third peak corresponds to scales in horizon 
crossing even before, being in the maximum of their third 
oscillation at the moment of the snapshot. 
The series of peaks continues at higher multipoles, 
with decreasing amplitude because of diffusion damping. 

As we mentioned in the Introduction, present data are strongly 
supporting this scenario, having revealed the first peak with 
very high confidence level and significant indications for 
a second and a third peak in the spectrum. 
Competing models for cosmological structure formation, 
see e.g. \cite{PZ} and references therein, 
predict markedly different spectra. Isocurvature models are generally 
characterized by a non-zero entropy perturbation between different 
species, keeping the curvature unperturbed; 
consequently, at horizon crossing 
the zero-point temperature fluctuation is zero, but not its 
time derivative, resulting in a sine time dependence instead 
of a cosine like in adiabatic models (\ref{dtt0adiaaini}), 
with a consequent shift of acoustic peaks 
by $\pi /2$ with respect to the adiabatic case. Coherent 
acoustic peaks are generally absent in non-Gaussian models 
like cosmological defects, because at horizon crossing 
both $(\delta T/T)_{0}$ and $(\delta T/T)_{1}$ can be different 
from zero, in a way which is different for each Fourier mode, 
thus destroying coherence. 

In the next Section we conclude this paper, giving a worked 
example on how the evidence for acoustic peaks in the CMB 
spectrum and their sensitivity to the main cosmological 
parameters can be used to constrain cosmological 
models. 

\begin{figure}[htb]
\vspace{9pt}
\includegraphics[width=15pc,height=8pc]{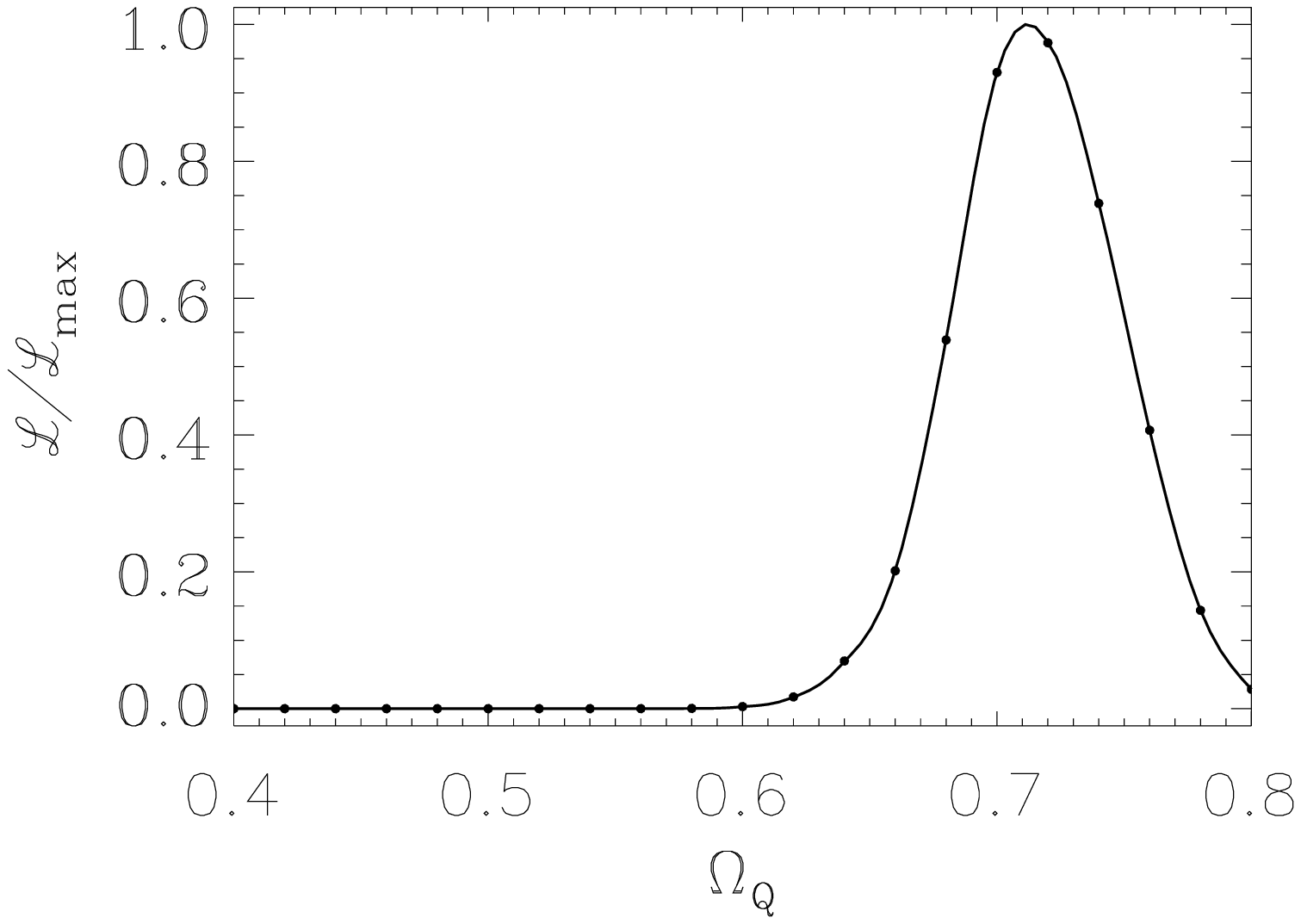}
\includegraphics[width=15pc,height=8pc]{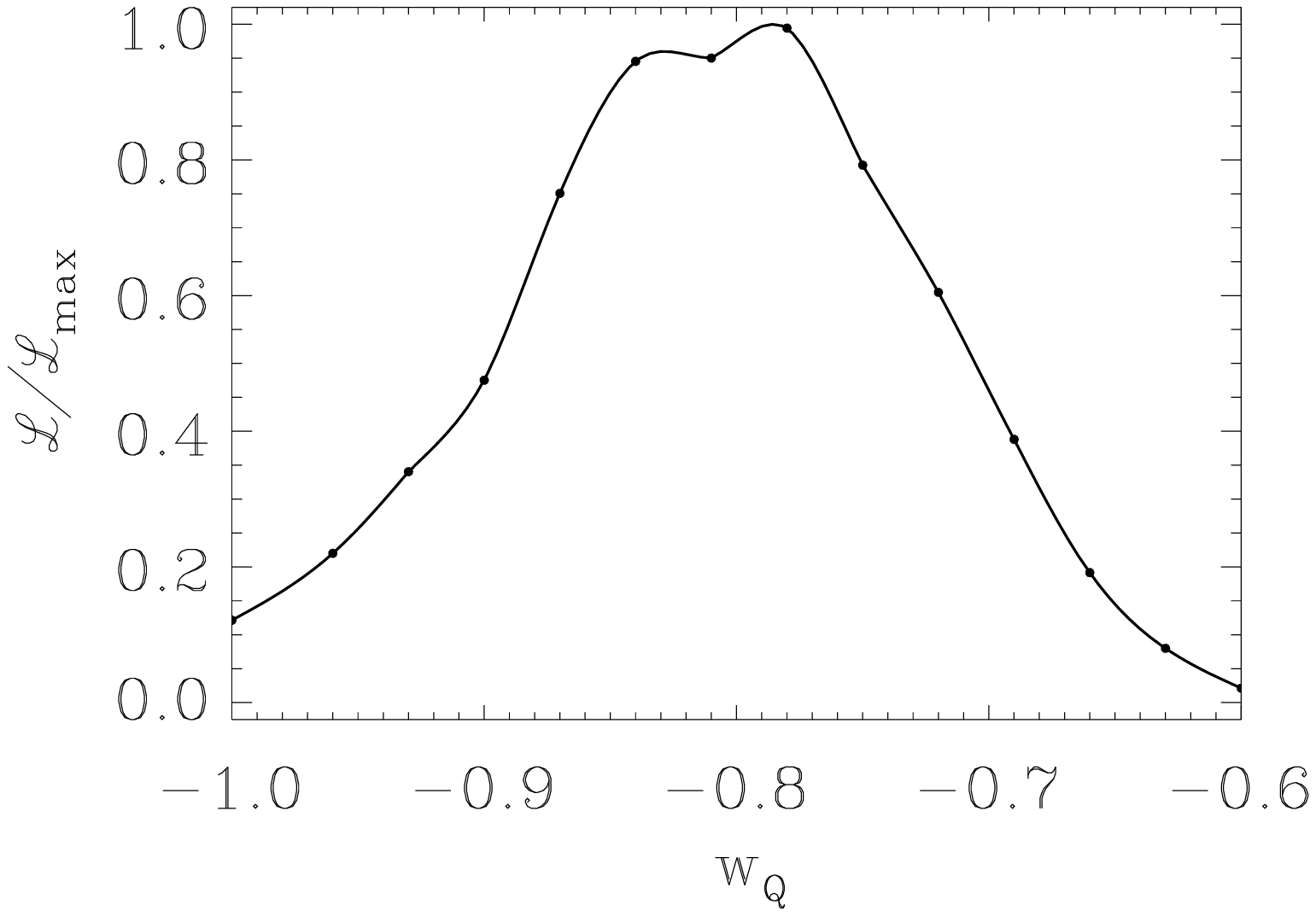}
\caption{Likelihood peaks for Quintessence parameters 
$\Omega_{Q}$ (top) and $w_{Q}$ (bottom) showing preference for Quintessence 
models with $w_{Q}\simeq -0.8$.}
\label{f3}
\end{figure}

\section{Constraining dark energy with CMB data}
\label{constraining}

As it is well known, CMB data are a powerful tool to 
constrain cosmological models, either because of 
the sensitivity on the most 
important cosmological parameters, either 
because they represent the Universe as it was 
before non-linear structure formation, thus being 
relatively simple to be read and understood. 
On the other hand, CMB alone cannot fix all cosmological 
parameters, either because of internal degeneracies \cite{EFSTA}, 
either because of the high number, about 10, of parameters 
to be investigated; independent 
datasets, mostly from Large Scale Structure (LSS) \cite{LSSQ}
and Type Ia supernovae \cite{SNQ} are needed in order to go 
deep in precision cosmology. 
However, under reasonable hypothesis, even the present 
CMB data, which of course do not reach the precision 
of the satellites MAP and Planck, 
can be used to derive interesting constraints on most 
important cosmological parameters. 

Here we give an example of this, constraining a 
dynamical vacuum energy in flat cosmologies with 
the present CMB data. 
The dark energy, also known as Quintessence, 
occupies a central position in modern cosmology after that 
at least three independent evidences, LSS, Supernovae and 
CMB, gave indications that almost $70\%$ of the cosmological 
energy density today resides in a sort of vacuum energy which 
is responsible for the cosmic acceleration today 
\cite{LSSQ,SNQ,BOOM,MAX,DASI}. To explain these 
observations a cosmological component with negative 
equation of state is necessary. Dark energy 
is described as a self-interacting scalar field 
rolling on its potential which admits dynamical 
trajectories, tracking solutions, in which 
its equation of state $w_{Q}$ can take values in 
the range [-0.5,-1], where the last value recovers 
the ordinary cosmological constant; 
these trajectories have been proved to exist both 
in ordinary and scalar-tensor cosmology 
(see e.g. \cite{TRACK} and references therein). 
The angle $\theta_{H}$ subtended by the horizon at 
decoupling is sensitive to the dark energy equation 
of state $w_{Q}$; indeed it is the ratio 
between the comoving value of Hubble horizon at decoupling, 
which is almost insensitive to values of $w_{Q}$ 
relevant for cosmic acceleration today, and the comoving 
distance of the last scattering surface from us, which is 
\begin{eqnarray}
\tau_{dec}\propto\int_{0}^{dec}
dz[\Omega_{m}(1+z)^{3}+\Omega_{K}(1+z)^{2}&+&\nonumber\\
+\Omega_{Q}(1+z)^{3(1+w_{Q})}]^{1/2}\ &,&
\label{taudec}
\end{eqnarray}
where $\Omega_{m}$, $\Omega_{K}$ and $\Omega_{Q}$ represent 
matter, curvature and Quintessence present energy density, 
respectively. 
Therefore the angle subtended by the horizon at decoupling is 
essentially inversely proportional to $\tau_{dec}$. 
As it is evident, the expression above is degenerate in the 
sense that a given $\tau_{dec}$ can be made of different 
combinations of the parameters entering into the integral. 
However, one should remember that this is not the only 
effect of dark energy on CMB; the change in the equation 
of state at low redshift enhance the CMB power on low 
multipoles $l\le 10$ and breaks the degeneracy between 
$\Omega_{Q}$ and $w_{Q}$ (see e.g. \cite{TRACK} and 
references therein). More serious is the degeneracy of 
dark energy with closed cosmological models with $\Omega_{K}>0$. 

In our recent work \cite{CMBQ} we fit CMB data 
\cite{BOOM,MAX,DASI} gridding several cosmological 
parameters as the baryon amount, cosmological 
gravitational waves and scalar spectral index, 
in addition to Quintessence parameters $\Omega_{Q}$ 
and $w_{Q}$, by assuming a number of priors including 
flatness $\Omega_{K}=0$. Interestingly, we find 
a preference of these data for dark energy models 
with respect to ordinary cosmological constant. 
This is shown in figure \ref{f3}, representing 
the likelihood for $\Omega_{Q}$, peaking at $70\%$, 
and equation of state, peaking at $w_{Q}\simeq -0.8$. 
This effect can be understood as follows: best fits 
obtained in the original works on the CMB data 
\cite{BOOM,MAX,DASI} slightly prefer 
closed cosmological models, even if flatness is well 
within errors. As it is evident from 
the expression (\ref{taudec}), 
dark energy models with $w_{Q}>-1$ induce a 
term which is similar to the one of closed models. 
Therefore, since we are assuming perfectly flat cosmologies, 
the best fit peak on dynamical vacuum energy models simply 
because they produce a similar geometrical effect on the 
angle subtended by the horizon at decoupling. 
Future data will help to test more deeply this 
interesting result. 

%


\begin{thebibliography}{9}
\bibitem{PZ} A. Liddle and D.H. Lyth (eds.), Cosmological Inflation
and Large Scale Structure, Cambridge University Press, 2000.
\bibitem{COBE} K.M. Gorski, Astrophys.J.S. 114 (1998) 1.
\bibitem{BOOM} C. B. Netterfield et al., submitted to Astrophys.J (2001), 
preprint astro-ph/0104460. 
\bibitem{MAX} A.T. Lee et al., submitted to Astrophys.J.Lett. (2001), 
preprint astro-ph/0104459. 
\bibitem{DASI} N.W. Halverson et al., submitted to Astrophys.J. (2001), 
preprint astro-ph/0104489
\bibitem{PY}P.J.E. Peebles and J.T. Yu, Astrophys.J. 162 (1970) 815. 
\bibitem{B}J.M. Bardeen, Phys.Rev.D22 (1980) 1882. 
\bibitem{KS}I. Kodama and M. Sasaki, Progr. of Theor.Phys.Supp, 78 (1984), 1.
\bibitem{H} J.C. Hwang, Astrophys.J. 375 (1991) 443. 
\bibitem{MB}C.P. Ma and E. Bertschinger Astrophys.J. 455 (1995) 7. 
\bibitem{HSWZ} W. Hu, U. Seljak, M. White, M. Zaldarriaga, 
Phys.Rev.D56 (1997) 596. 
\bibitem{EFSTA} G. Efstathiou, submitted to MNRAS (2001), 
preprint astro-ph/0109151. 
\bibitem{LSSQ} J.A. Peacock et at., Nature 410 (2001) 169
\bibitem{SNQ} S. Perlmutter et at., Astrophys.J. 517 (1999) 565; 
A. Riess et al., Astron.J. 116 (1998) 1009. 
\bibitem{TRACK} C. Baccigalupi, S. Matarrese, F. Perrotta 
Phys.Rev.D62 (2000) 123510. 
\bibitem{CMBQ} C. Baccigalupi, A. Balbi, S. Matarrese, F. Perrotta, 
N. Vittorio, submitted to Phys.Rev.D (2001), preprint astro-ph/0109097. 

\end{thebibliography}
\end{document}